\documentclass[12pt,a4paper]{article}
\usepackage[english]{babel}
\usepackage{epsfig}

\begin{document}

\begin{center}
{\Large \bf A discrete version of the inverse scattering problem
and the J-matrix method}\\

{\large S.~A.~Zaytsev\footnote{This work has been done partially
while the author was visiting the Institute for Nuclear Theory,
University of Washington.}}\\

{\sl Department of Physics, Khabarovsk State University of Technology,\\
Tikhookeanskaya 136, Khabarovsk 680035, Russia}\\[4pt]
\end{center}
\begin{abstract}
The problem of the Hamiltonian matrix in the oscillator and
orthogonalized Laguerre basis construction from a given S-matrix
is treated in the context of the algebraic analogue of the
Marchenko method.
\end{abstract}

\section{Introduction}
The J-matrix \cite{HY} theory of scattering is based on the fact
that the $\ell$th partial wave kinetic energy or the Coulomb
Hamiltonian $H^0$ is represented in a certain square-integrable
basis set by an infinite symmetric tridiagonal matrix. In the
harmonic oscillator and the Laguerre basis sets
$\left\{\phi_n^{\ell}\right\}_{n=0}^{\infty}$ the eigenvalue
problem for $H^0$ can be solved analytically. The J-matrix method
yields an exact solution to a model scattering Hamiltonian where
the given short-range potential is approximated by truncating in a
finite subset $\left\{\phi_n^{\ell}\right\}_{n=0}^{N-1}$.

In Refs. \cite{Z1,Z2} an inverse scattering formalism within the
J-matrix method has been proposed, where the matrix $\|V_{n,m}\|$
of the potential
\begin{equation}
V_{\ell}(r, \, r')=\hbar\omega \sum \limits_{n, \, m=0}^{N-1}
\phi_n^{\ell}(x)\,V_{n, \, m}\,\phi_m^{\ell}(x') \label{Pot}
\end{equation}
with the oscillator form factors
\begin{equation}
\phi_n^{\ell}(x)=(-1)^n\,\sqrt{\frac{2n!}{\rho
\Gamma(n+\ell+\frac32)}}\,
x^{\ell+1}e^{-x^2/2}\,L_n^{\ell+1/2}(x^2) \label{bf}
\end{equation}
is determined from a given S-matrix. Here, $x=r/\rho$ is the
relative coordinate in units of the oscillator radius
$\rho=\sqrt{\hbar/\mu\omega}$, $\mu$ is the reduced mass.

Obviously a correlation can be made between the J-matrix method
and a discrete model of quantum mechanics, within of which a
finite-difference Schr\"{o}dinger equation is used. As a result
the J-matrix versions of the Gel'fand-Levitan-Marchenko method
algebraic analogue can be formulated. For instance, the J-matrix
method formally and computationally is quite similar to the
R-matrix theory. It is this analogy that the previous J-matrix
version of inverse scattering theory \cite{Z1,Z2,ZK} [also see
\cite{PC}] leans upon. Within the J-matrix approach the discrete
representation of the Green function in finite subspace of the
basis functions $\left\{ \phi_n^{\ell}\right\}_{n=0}^{N-1}$
\begin{equation}
{\bf G}(\epsilon)=\left(\epsilon {\bf I}-{\bf h}\right)^{-1}
\end{equation}
is used. Here, ${\bf I}$ is identity matrix and ${\bf h}$ is the
truncated Hamiltonian matrix of order $N$ in the oscillator basis
(\ref{bf}). We measure the energy $E$ in the units of the
oscillator basis parameter $\hbar\omega$, i.e.
$E=\hbar\omega\epsilon$ and $\epsilon=q^2/2$, where $q$ is the
dimensionless  momentum: $q=k\rho$. In particular, the element
$\left[{\bf G}(\epsilon)
\right]_{N-1,\,N-1}\equiv\mathcal{P}_N(\epsilon)$ can be
presented in the two rational forms \cite{Z1,YAA}:
\begin{equation}\label{GR1}
  \mathcal{P}_N(\epsilon)= \frac{\prod \limits _{j=0}^{N-2}(\epsilon-\mu_j)}
  {\prod \limits_{j=0}^{N-1}(\epsilon-\lambda_j)}
\end{equation}
where $\left\{\lambda_j\right\}_{j=0}^{N-1}$ and
$\left\{\mu_j\right\}_{j=0}^{N-2}$ satisfy the interlacing
property
$$
\lambda_0 < \mu_0 < \lambda_1 < \ldots < \lambda_{N-2} < \mu_{N-2} < \lambda_{N-1},
$$
and \cite{YF}
\begin{equation}\label{GR2}
  \mathcal{P}_N(\epsilon)= \sum \limits_{j=0}^{N-1}\frac{Z_{N-1,\,j}^2}
  {\epsilon-\lambda_j}.
\end{equation}
Here, $\left\{Z_{N-1,\,j} \right\}_{j=0}^{N-1}$ are the elements
of the $N$th row of the eigenvector matrix ${\bf Z}$ of the
truncated Hamiltonian matrix ${\bf h}$. The sets
$\left\{\lambda_j\right\}_{j=0}^{N-1}$,
$\left\{\mu_j\right\}_{j=0}^{N-2}$ and
$\left\{\lambda_j\right\}_{j=0}^{N-1}$, $\left\{Z_{N-1,\,j}
\right\}_{j=0}^{N-1}$ are derived from the S-matrix which is
intimately connected with $\mathcal{P}_N$ [see Eq. (\ref{SN})].
Notice that the both sets of the spectral parameters determine
unique [apart from the off diagonal elements sign] Hamiltonian
matrix ${\bf h}$ of a Jacobi form \cite{G,GW}
\begin{equation}
\|h_{n,\,m}\| = \left(
 \begin{array}{cccccc}
  a_0 & b_1 & & &   \\
  b_1 & a_1 & b_2 & & \mbox{\Large $0$} &  \\
      & b_2 & a_2 & b_3 & &  \\
  && b_3& \times & \times &  \\
  &\mbox{\Large $0$} &&\times & \times & b_{N-1}\\
  &&&& b_{N-1} & a_{N-1} \\
 \end{array}
\right). \label{hnm}
\end{equation}
Hence, the sought-for potential matrix $\|V_{n,\,m}\|$ is also of
a Jacobi form. Recall that the kinetic energy operator
\begin{equation}
H^0 = \frac{\hbar^2}{2\mu}\left(-\frac{\displaystyle
d^2}{\displaystyle d\,r^2}+ \frac{\ell(\ell+1)}{\displaystyle
r^2}\right)\label{H0}
\end{equation}
matrix representation
$\|T_{n,\,m}^{\ell}\|=\frac{1}{\hbar\omega}\|H^0_{n,\,m}\|$ in the
harmonic oscillator basis (\ref{bf}) is symmetric tridiagonal
\cite{HY,YF}:
\begin{equation}
 \begin{array}{c}
T_{n,\,n}^{\ell}=\frac12\left(2n+\ell+\frac32\right),\\[3mm]
T_{n,\,n+1}^{\ell}=T_{n+1,\,n}^{\ell}=-\frac12\sqrt{(n+1)
\left(n+\ell+\frac32\right)}.\\
 \end{array}
 \label{T}
\end{equation}

Thus the inverse scattering problem within the J-matrix approach
admits of the solution in the tridiagonal Hamiltonian matrix
form. In this regard the J-matrix method is similar to a discrete
model of quantum mechanics, in the framework of which the
Hamiltonian matrix representation is also symmetric tridiagonal.
Note that the tridiagonal matrix representation of both the
kinetic energy operator and the Hamiltonian is fundamental for a
finite-difference analogue of the Gel'fand-Levitan equations [see
e.g. \cite{Chabanov}] as well as for a discrete version
\cite{Case} of the Marchenko equations. As shown below, the
similarity between the J-matrix method and a finite-difference
approach can be plainly extended to the inverse scattering
formalism as well. In the present paper, in particular, an
inverse scattering J-matrix approach via the Marchenko equations
(JME) is given.

In JME the expansion coefficients $c_n$ of the wave function
$\psi$ in terms of the $L^2$ basis set play a role similar to
that of the values $\psi_n=\psi(x_n)$ at points $x_n=n\Delta$
within the finite-difference inverse scattering approach. Here,
the completeness relation for the solutions $c_n$ of the
Schr\"{o}dinger equation discrete analogue is also exploited which
involves an integration $c_n\overline{c_m}$ over the energy from
zero to infinity. This raises the question as to whether the
integrals converge. As shown below, taking account of the phase
shift $\delta^N_{\ell}(k)$ corresponding to the potential
(\ref{Pot}) of finite rank $N$ asymptotic behavior at large $k$
provides the convergence of the integrals. Generally, with a
potential (\ref{Pot}) matrix of finite order $N$ it is possible to
reproduce the phase shift $\delta_{\ell}$ only on a finite energy
interval $[0, \epsilon_0]$ with $\epsilon_0 < \lambda_{N-1}$.
This is why the eigenvalue $\lambda_{N-1}$ and the corresponding
eigenvector component $Z_{N-1, \,N-1}$ are the variational
parameters within the previous J-matrix version
\cite{Z1,Z2,ZK,PC} of the inverse scattering theory. By contrast,
in JME the phase shift, even if modified in accordance with the
$\delta_{\ell}^N$ asymptotic feature, on infinite energy interval
is used. As a result JME has not variational parameters (apart
from $N$ and $\rho$).

The elements of the J-matrix method formalism are presented in
Sec.~2. In Sec.~3 the inverse scattering J-matrix approach in the
context of the Marchenko equations is formulated. The features of
JME numerical realization are discussed in Sec.~4. In Sec.~5 JME
is expanded to the Laguerre basis case. Here, we are dealing with
the tridiagonal Hamiltonian matrix construction in an
orthogonalized Laguerre basis set, in which the kinetic energy
operator matrix is also tridiagonal. In Sec.~6 we summarize our
conclusions.

\section{The direct problem}
The oscillator-basis J-matrix formalism is discussed in detail
elsewhere. We present here only some relations needed for
understanding the inverse scattering J-matrix approach. Within
the J-matrix method, the radial wave function $\psi(k,\,r)$ is
expanded in an oscillator function (\ref{bf}) series
\begin{equation}
\psi(k,\,r) = \sum \limits_{n=0}^{\infty}c_n(k) \,
\phi_n^{\ell}(x). \label{rk}
\end{equation}
In the assumption that the Hamiltonian matrix is of the form
(\ref{hnm}) the functions $c_n$ are the solutions to the set of
equations
\begin{equation}
 \begin{array}{c}
 a_0\,c_0(k)+b_1\,c_1(k)=\epsilon\,c_0(k)\\[3mm]
 b_n\,c_{n-1}(k)+a_n\,c_n(k)+b_{n+1}\,c_{n+1}(k)=\epsilon\,c_n(k),
 \quad n=1,\, 2, \, \ldots \, .\\
 \end{array}
\label{PEq}
\end{equation}
The asymptotic behavior of $c_n(k)$ for $k>0$ as $n \rightarrow
\infty$ is given by
\begin{equation}
c_n(k)=f_n(k)\equiv \frac{\mbox{i}}{2}
\left[\;\mathcal{C}_{n,\,\ell}^{(-)}(q)-S(k)\,\mathcal{C}_{n,\,\ell}^{(+)}(q)\;
\right]. \label{f01}
\end{equation}
Here, the functions
\begin{equation}
 \begin{array}{c}
\mathcal{C}_{n,\,\ell}^{(\pm)}(q)=\mathcal{C}_{n,\,\ell}(q)
\pm\mbox{i}\mathcal{S}_{n,\,\ell}(q),\\[3mm]
\mathcal{S}_{n,\,\ell}(q) = \sqrt{\frac{\pi\rho
n!}{\Gamma(n+\ell+\frac32)}}\,
q^{\ell+1}e^{-q^2/2}\,L_n^{\ell+1/2}(q^2),\\[3mm]
\mathcal{C}_{n,\,\ell}(q) = \sqrt{\frac{\pi\rho
n!}{\Gamma(n+\ell+\frac32)}}\,
\frac{\Gamma(\ell+1/2)}{\pi\,q^{\ell}}\,e^{-q^2/2}\,
F\left(-n-\ell-1/2,\, -\ell+1/2;\, q^2 \right)\\
 \end{array}
 \label{SC}
\end{equation}
obey the ``free'' equations
\begin{equation}
T_{n, \, n-1}^{\ell}\,e_{n-1}(q)+ T_{n, \,
n}^{\ell}\,e_n(q)+T_{n, \, n+1}^{\ell}\,e_{n+1}(q)=\epsilon
\,d_n(q), \quad n=1, \, 2, \, \ldots \, .\label{TRR}
\end{equation}
$\mathcal{S}_{n,\,\ell}$ satisfy in addition the equation
\begin{equation}
T_{0, \, 0}^{\ell}\,\mathcal{S}_{0,\,\ell}(q)+T_{0, \,
1}^{\ell}\,\,\mathcal{S}_{1,\,\ell}(q) =
\epsilon\,\mathcal{S}_{0,\,\ell}(q).
\end{equation}
Besides, $\mathcal{S}_{n,\,\ell}$ meet the completeness relation
\begin{equation}
\frac{2}{\pi}\int \limits_0^{\infty}dk\,\mathcal{S}_{n,\,\ell}(q)
\,\mathcal{S}_{m,\,\ell}(q)=\delta_{n,\,m}. \label{SCompl}
\end{equation}

Notice that
$$
\begin{array}{c}
\widetilde{S}(r)=\sum \limits
_{n=0}^{\infty}\mathcal{S}_{n,\,\ell}(q)\,\phi_n^{\ell}(x),\\[3mm]
\widetilde{C}(r)=\sum \limits
_{n=0}^{\infty}\mathcal{C}_{n,\,\ell}(q)\,\phi_n^{\ell}(x),\\
\end{array}
$$
subject to the asymptotic condition  \cite{YF}
\begin{equation}\label{scasym}
\begin{array}{c}
\widetilde{S}(r) \mathop{\sim}\limits_{r\rightarrow\infty}\sin(kr-\ell\pi/2),\\[3mm]
\widetilde{C}(r) \mathop{\sim}\limits_{r\rightarrow\infty}\cos(kr-\ell\pi/2).\\
\end{array}
\end{equation}

As for the coefficients of the expansion
\begin{equation}
\psi_{\nu}(r) = \sum
\limits_{n=0}^{\infty}c_n(\mbox{i}\kappa_{\nu}) \,
\phi_n^{\ell}(x) \label{ik}
\end{equation}
of the normalized bound state wave function $\psi_{\nu}$ with the
energy $-\kappa_{\nu}^2/2$,
\begin{equation}
c_n(\mbox{i}\kappa_{\nu})=f_n(\mbox{i} \kappa_{\nu})\equiv
\mathcal{M}_{\nu}\,
\mbox{i}^{\ell}\,\mathcal{C}_{n,\,\ell}^{(+)}(\mbox{i}\kappa_{\nu}\rho)
\label{f02}
\end{equation}
holds as $n \rightarrow \infty$. Here, $\mathcal{M}_{\nu}$ is the
bound state normalization constant which is related to the
residue of the S-matrix \cite{Baz}:
\begin{equation}
\mbox{i}\mathop{Res}\limits _{k=\mbox{\scriptsize
i}\kappa_{\nu}}S(k) = (-1)^{\ell}\, \mathcal{M}_{\nu}^2.
\label{M}
\end{equation}

It can be easy verified that from the completeness relation for
the solutions $\psi(k, \, r)$, $\psi_{\nu}(r)$ \cite{CS}
\begin{equation}
\frac{2}{\pi}\int \limits _0^{\infty} dk
\,\psi(k,\,x)\,\overline{\psi(k,\,y)} + \sum \limits _{\nu}
\psi_{\nu}(x)\, \overline{\psi_{\nu}(y)}= \delta(x-y)
\label{ComplSE}
\end{equation}
it follows that
\begin{equation}
\frac{2}{\pi}\int \limits _0^{\infty} dk \,
c_n(k)\,\overline{c_m(k)} + \sum \limits _{\nu}
c_n(\mbox{i}\kappa_{\nu})\, \overline{c_m(\mbox{i}\kappa_{\nu})}=
\delta_{n,\, m}. \label{Compl}
\end{equation}

\section{The inverse problem}
To take advantage of the algebraic analogue of the Marchenko
method it is essential that there exist coefficients $K_{n, \,
m}$ [independent of $k$] such that
\begin{equation}
c_n(k) = \sum \limits _{m=n}^{\infty}K_{n, \, m}\, f_m(k).
\label{EC}
\end{equation}
By analogy with Ref. \cite{Case} assume that
\begin{equation}
c_n(k) = \, f_n(k), \quad n \ge N
\end{equation}
[$N$ specifies the order of a potential matrix]. If $f_N$ and
$f_{N+1}$ are inserted [instead of respectively $c_N$ and
$c_{N+1}$] into Eq. (\ref{PEq}) for $n=N$, we obtain, in view of
Eq. (\ref{TRR}),
\begin{equation}
 \begin{array}{c}
c_{N-1}(k) = \left( \right.T_{N,\, N-1}^{\ell}f_{N-1}(k)+[T_{N,\,
N}^{\ell}-a_N]\,f_{N}(k)+\qquad \qquad\\[3mm]
\qquad \qquad \qquad \qquad \qquad \qquad
+[T_{N,\,N+1}^{\ell}-b_{N+1}]\,f_{N+1}(k)]
\left.\right)/b_{N}.\\
 \end{array}
\label{fN}
\end{equation}
Then, using the three-term recursion relation (\ref{TRR}) with
every $n=N~-1, \ldots,1$ we obtain
\begin{equation}
c_n(k) = \sum \limits _{m=n}^{2N-n-1}K_{n, \, m}\, f_m(k), \quad
n=0, \, 1, \, \ldots, \, N-1 \label{Sol}
\end{equation}
[which in the limit $N \rightarrow \infty$ gives (\ref{EC})].

The coefficients $K_{n,\, m}$ are found from the completeness
relation (\ref{Compl}). From the condition of the orthogonality
of $c_n$ and every $c_m$, $m > n$ follows the condition of the
orthogonality of $c_n$ and every $f_m$, $m > n$, i. e.
\begin{equation}
\frac{2}{\pi}\int \limits _0^{\infty} dk \,
c_n(k)\,\overline{f_m(k)} + \sum \limits _{\nu}
c_n(\mbox{i}\kappa_{\nu})\,\overline{f_m(\mbox{i}\kappa_{\nu})}=
0, \quad m > n. \label{Em}
\end{equation}
Inserting the expansion of (\ref{Sol}) in (\ref{Em}) gives the
system of linear equations in $K_{n, \, m}$
\begin{equation}
 K_{n,\, n}\,Q_{n, \,m}+\sum \limits_{p=n+1}^{2N-n-1} K_{n, \,
 p}\,Q_{p, \, m}=0, \quad m>n.
 \label{KEq1}
\end{equation}
Then, inserting Eq. (\ref{Sol}) into (\ref{Compl}) and putting
$n=m$, we obtain, in view of Eq. (\ref{Em}), the equation in
$K_{n,\,n}$
\begin{equation}
 K_{n,\, n}\left(K_{n,\, n}\,Q_{n,\, n}+\sum \limits_{p=n+1}^{2N-n-1} K_{n, \,
 p}\,Q_{p, \, n} \right)=1.
 \label{KEq2}
\end{equation}
Note that from Eq. (\ref{KEq1}) it follows that $K_{n, \, m}$,
$m>n$ are proportional to $K_{n,\,n}$. In Eqs. (\ref{KEq1}) and
(\ref{KEq2}) $Q_{n,\, m}$ are defined from the scattering data by
\begin{equation}
Q_{n, \, m}= \frac{2}{\pi}\int \limits _0^{\infty} dk \,
f_n(k)\,\overline{f_m(k)} + \sum \limits _{\nu}
f_n(\mbox{i}\kappa_{\nu})\, \overline{f_m(\mbox{i}\kappa_{\nu})}.
\label{Q}
\end{equation}

The elements $a_n$ and $b_n$ of the sought-for Hamiltonian matrix
(\ref{hnm}) are related to $K_{n,\,m}$ by the equations
\begin{equation}
 \begin{array}{c}
 a_n = T_{n,\,n}^{\ell}+\frac{K_{n,\,n+1}}{K_{n,\,n}}\,T_{n+1,\,n}^{\ell}-
\frac{K_{n-1,\,n}}{K_{n-1,\,n-1}}\,T_{n,\,n-1}^{\ell}, \\[3mm]
b_n=\frac{K_{n,\,n}}{K_{n-1,\,n-1}}\,T_{n,\,n-1}^{\ell}, \quad
n=1,\, 2, \, 3, \, \ldots.\\
 \end{array}
 \label{abK}
\end{equation}
$a_0$ is specified by the solutions $c_0$ and $c_1$ to Eq.
(\ref{PEq}).

\section{A numerical realization}
To this point the assumption has been made that the phase shift
$\delta_{\ell}(k)$ is a continuous function of the wave number
$k$ that meets the conditions \cite{CS}
$$
\delta_{\ell}(\infty)=0, \qquad \int \limits
^{\infty}k^{-1}|\delta_{\ell}(k)|dr < \infty.
$$
In this case $\delta_{\ell}$ must satisfy stringent requirements.
Indeed, the integrated function in r.h.s. of Eq.(\ref{Q}) can be
expressed in the form of a product $g_n\,g_m$ of (real) functions
\begin{equation}\label{fn}
  g_n(q)= \mathcal{S}_{n\,\ell}(q)\,\cos\delta_{\ell}
  +\mathcal{C}_{n\,\ell}(q)\,\sin\delta_{\ell}.
\end{equation}
It is obviously that a sufficient condition to the convergence of
the integrals in (\ref{Q}) is that functions $g_n$ are
square-integrable. Notice that from (\ref{SC}) follows
\begin{equation}\label{AS}
\mathcal{S}_{n\,\ell}(q)
\mathop{\sim}\limits_{q\rightarrow\infty} (-1)^n
\sqrt{\frac{\pi\rho}{n!\;\Gamma(n+\ell+\frac32)}}\,q^{2n+\ell+1}\,e^{-q^2/2},
\end{equation}
i.e. the first term in (\ref{fn}) decays exponentially at
asymptotically large $q$. However, $\mathcal{C}_{n\,\ell}$ grows
exponentially with increasing $q$:
\begin{equation}\label{AC}
  \mathcal{C}_{n\,\ell}(q)\mathop{\sim}\limits_{q\rightarrow\infty}
 (-1)^{n+1}\frac{\sqrt{\pi\rho\,n!\;\Gamma(n+\ell+\frac32)}}{\pi}\,
  q^{-(2n+\ell+2)}e^{q^2/2}.
\end{equation}
This suggests that the phase shift $\delta_{\ell}$ must decay
rapidly enough to provide the convergence of the integral in
r.h.s. of Eq. (\ref{Q}).

Actually the phase shift $\delta_{\ell}^N$ corresponding to the
potential (\ref{Pot}) of rank $N$ \cite{YF}
\begin{equation}\label{tdl}
  \tan \delta_{\ell}^N=-\frac{\mathcal{S}_{N-1,\, \ell}(q)-
  \mathcal{P}_N(\epsilon)\,T_{N-1,\, N}^{\ell}\,\mathcal{S}_{N,\, \ell}(q)}
  {\mathcal{C}_{N-1,\, \ell}(q)-\mathcal{P}_N(\epsilon)\,T_{N-1,\,
  N}^{\ell}\,
  \mathcal{C}_{N,\, \ell}(q)},
\end{equation}
as seen in Eqs. (\ref{AS}), (\ref{AC}), fulfills even more strict
requirement
\begin{equation}\label{dla}
  \delta_{\ell}^N\mathop{\sim}
  \limits_{q\rightarrow\infty}
   \frac{\pi\,(2N+\ell-\frac12)}{(N-1)!\;\Gamma(N+\ell+\frac12)}\,
   q^{4N+2\ell-3}\,e^{-q^2}.
\end{equation}
Because of the restriction (\ref{dla}) on the phase shift
$\delta_{\ell}^N$ the potential (\ref{Pot}) of finite rank $N$
generally is incapable to describe the scattering data on the
infinite interval $k \in [0, \, \infty)$. At most, we can set
ourselves the task of constructing the potential (\ref{Pot}) that
describes the experimental phase shift $\delta_{\ell}$ on some
finite interval $[0, \, k_0]$, since generally $\delta_{\ell}$
needs to be modified in the region $k>k_0$ to provide at least the
convergence of the integrals in Eq. (\ref{Q}).

As an example we consider the $s$-wave scattering case. The
``experimental'' phase shift $\delta_{\ell}$ [dotted curve in
figure~1] is that of the scattering on the potential given by
straight well with the depth $V_0$:
$\sqrt{\frac{2\mu}{\hbar^2}\,V_0}\,R=1.5$. The potential
(\ref{Pot}) is sought for that describes the phase shift on the
interval $[0, \, k_0]$, $k_0R=6$ [in figure~1 crosses represent a
modified phase shift]. The phase shift $\delta_{\ell}^{N}$
corresponding to the resulting potential (\ref{Pot}) of rank
$N=6$ and $\rho=\frac12R$ is shown in figure~1 [solid curve].
Notice that $\mathcal{C}_{n\,\ell}$ explodes exponentially with
increasing $q$. Thus, the contribution from the region
$q>q_0=\rho k_0=3$ to the integral in Eq. (\ref{Q}) may become
overwhelming [see figure~2 where $g_0(q)^2$ is plotted], with the
result that the method fails. Matters can be improved by a
transition to lesser $\rho$ that shifts $q_0$ to a region where
$\mathcal{C}_{n\,\ell}$ is not that large, or replacing
$\delta_{\ell}$ at $q>q_0$ with a function that decays rapidly
enough.

In the second example, a scattering data on the potential with a
bound state has been used as input. The phase shift
$\delta_{\ell}$ [dotted curve in figure~3] corresponds to the
s-wave scattering on a spherically symmetric potential in the
form of straight well. The well parameter
$\sqrt{\frac{2\mu}{\hbar^2}\,V_0}\,R=2$ determines the bound
state with the energy $E=-\kappa^2$, $\kappa R=0.638045$ and the
asymptotic normalization constant $\mathcal{M}R^{1/2}=1.583324$.
$k_0$ and $\rho$ have been taken the same as in the first example
[the modified phase shift is represented by crosses in figure~3].
It is well known that a phase shift does not depend on energy
positions and asymptotic normalization constants of bound states.
Thus, the inverse scattering problem in the presence of a bound
state can be split into two steps.

On the first step we focus on the describing the phase shift and,
in spite of the whole of scattering data is used [see
Eq.(\ref{Q})], do not seek to describe the bound state with high
degree of accuracy. The phase shift $\delta_{\ell}^{N}$
corresponding to the potential (\ref{Pot}) parameters, which
together with $\kappa R$ and $\mathcal{M}R^{1/2}$ are presented
in the left half of Table, is shown in figure~3 [solid curve].

On the second step, to improve the description of the bound
states we use the relationship (\ref{M}) between the poles and
residues of the S-matrix and the characteristics of the bound
states [see e.g. Ref. \cite{PC}]. Here, the smallest eigenvalue
$\lambda_0$ and the corresponding eigenvector component
$Z_{N-1,\,0}$ associated with the bound state are found from the
system
\begin{equation}\label{ZN}
\sum \limits _{j=0}^{N-1} \,Z_{N-1,\,j}^2=1,
\end{equation}
\begin{equation}\label{DZ}
  \mathcal{P}_N(-\kappa^2\rho^2/2)
    =\frac{1}{T_{N-1,\, N}^{\ell}}\frac{\mathcal{C}_{N-1,\, \ell}^{(+)}
    (\mbox{i}\kappa\rho)}
  {\mathcal{C}_{N,\, \ell}^{(+)}(\mbox{i}\kappa\rho)},
\end{equation}
\begin{equation}\label{DDZ}
\frac{\mathcal{C}_{N-1,\, \ell}^{(-)}(\mbox{i}\kappa\rho)-
  \mathcal{P}_N(-\kappa^2 \rho^2/2)\,T_{N-1,\, N}^{\ell}\,
  \mathcal{C}_{N,\, \ell}^{(-)}(\mbox{i}\kappa\rho)}
  {\frac{d}{dq}\left\{\mathcal{C}_{N-1,\, \ell}^{(+)}(q)-
  \mathcal{P}_N(q^2/2)\,T_{N-1,\,
  N}^{\ell}\,
  \mathcal{C}_{N,\, \ell}^{(+)}(q)\right\}\left.\vphantom{I^I}
   \right|_{q=\mbox{\scriptsize i}\kappa\rho}}  =\mbox{i}
   (-1)^{\ell+1}\, \rho \mathcal{M}^2.
\end{equation}
Eqs. (\ref{DZ}), (\ref{DDZ}) are derived from the S-matrix
formula for the potential (\ref{Pot}) \cite{YF}
\begin{equation}\label{SN}
  S_{\ell}^N=\frac{\mathcal{C}_{N-1,\, \ell}^{(-)}(q)-
  \mathcal{P}_N(\epsilon)\,T_{N-1,\, N}^{\ell}\,\mathcal{C}_{N,\, \ell}^{(-)}(q)}
  {\mathcal{C}_{N-1,\, \ell}^{(+)}(q)-\mathcal{P}_N(\epsilon)\,T_{N-1,\,
  N}^{\ell}\,
  \mathcal{C}_{N,\, \ell}^{(+)}(q)}
\end{equation}
and Eq. (\ref{M}). Notice that the component $Z_{N-1,\,N-1}$
corresponding to the leading eigenvalue $\lambda_{N-1}$ is
involved to meet the normalization condition (\ref{ZN}). The phase
shift is scarcely affected by changing the parameters
$\left\{\lambda_0, \, Z_{N-1, \, 0}\right.$, $\left.Z_{N-1, \,
N-1} \right\}$ from the initial values obtained on the first step
to the ones that are evaluated from Eqs.(\ref{ZN})~-~(\ref{DDZ}).
The potential parameters, which provide the correct values of
$\kappa R$ and $\mathcal{M}R^{1/2}$, are presented in the right
half of Table.

\section{The Laguerre basis}
\subsection{Preliminaries}
For simplicity's sake we restrict our consideration to the
scattering of neutral particles. However, the resulting equations
still stand in the presence of the repulsive Coulomb interaction.
The potential sought is given by the expression
\begin{equation}
V_{\ell}(r, \, r')=\frac{\hbar^2}{2\mu} \sum_{n, \, m =0}^{N-1}
\overline{\phi}_n^{\ell}(x) V_{n, \,
m}\overline{\phi}_{m}^{\ell}(x') \label{LPot}
\end{equation}
where the functions $\overline{\phi}_n^{\ell}$
\begin{equation}
\overline{\phi}_n^{\ell}(x) =\frac{\displaystyle n!}
  {\displaystyle r \, (n+2\ell+1)!}\, \phi_n^{\ell}(x) \label{OLbf}
\end{equation}
are bi-orthogonal to the base Laguerre functions $\phi_n^{\ell}$:
\begin{equation}
\phi_n^{\ell}(x) = (2 b r)^{\ell+1} \,
  e^{-b r} L_n^{2\ell+1}(2 b r),\label{Lbf}
\end{equation}
i.e.
\begin{equation}
\int \limits_0^{\infty} dr \overline{\phi}_n^{\ell}(x)\,
\phi_m^{\ell}(x)=\delta_{n,\,m}. \label{O1}
\end{equation}
Here, $b$ is the scale parameter: $x=br$.

The coefficients $u_n$ of the expansion
\begin{equation}
\psi(k,\,r) = \sum \limits_{n=0}^{\infty}u_n(k) \,
\phi_n^{\ell}(x) \label{Lrk}
\end{equation}
of the Schr\"{o}dinger equation regular solution $\psi(k,\,r)$
satisfy the system of equations
\begin{equation}
\left(h^0_{n, \, m}+V_{n, \, m} \right)u_m(k)=k^2\,A^{\ell}_{n, \,
m}\,u_m(k), \qquad n=0,\, 1,\, \ldots\,  . \label{LEq}
\end{equation}
Here, $\|h^0_{n, \, m}\|$ is the symmetric tridiagonal matrix of
the reference Hamiltonian $\frac{2\mu}{\hbar^2}H^0$ (\ref{H0})
calculated in the basis (\ref{Lbf}) \cite{HY}:
\begin{equation}
 \begin{array}{c}
h^0_{n,\,n}=
b\frac{(n+2\ell+1)!}{n!}(n+\ell+1),\\[3mm]
h^0_{n,\,n+1}=h^0_{n+1,\,n}=
b\frac{(n+2\ell+2)!}{2n!}, \qquad n=0,\, 1, \, \ldots \,.\\
 \end{array}
\label{LH0}
\end{equation}
$\|A^{\ell}_{n, \, m}\|$ signifies the basis-overlap matrix
\begin{equation}
A^{\ell}_{n, \, m}=\int \limits_0^{\infty} dr
\phi_n^{\ell}(x)\,\phi_m^{\ell}(x)\label{OVLD}
\end{equation}
which is also of Jacobi form:
\begin{equation}
 \begin{array}{c}
A^{\ell}_{n,\,n}=\frac{(n+2\ell+1)!}{b\,n!}(n+\ell+1),\\[3mm]
A^{\ell}_{n,\,n+1}=A_{n+1,\,n}=-\frac{(n+2\ell+2)!}{2bn!},
\qquad n=0,\, 1, \, \ldots \,.\\
 \end{array}
\label{OVLM1}
\end{equation}

The asymptotic behaviour of the coefficients  $u_n(k)$, $k>0$, as
$n\rightarrow\infty$ is given by the following expression:
\begin{equation}\label{Lkp}
u_n(k)=w_n(k) \equiv \frac{\mbox{
i}}{2}\left[\mathcal{C}_{n,\,\ell}^{(-)}(k)-S(k)\,\mathcal{C}_{n,\,\ell}^{(+)}(k)
\right]
\end{equation}
where the functions \cite{YF}
\begin{equation}
 \begin{array}{c}
\mathcal{C}_{n,\,\ell}^{(\pm)}(k)=-\frac{n!}{(n+\ell+1)!}
\frac{(-\xi)^{\pm(n+1)}}{\left(2\, \sin \zeta \right)^{\ell}}\,
{_2F_1}(-\ell, \, n+1; \; n+\ell+2; \; \xi^{\pm 2}),\\[3mm]
\xi=e^{\mbox{\scriptsize i}\zeta}=\frac{ \displaystyle\mbox{i}b -
k }
{ \displaystyle \mbox{i}b + k },\\
 \end{array}
\label{LCpm}
\end{equation}
obey the inhomogeneous ``free'' equation
\begin{equation}
J^{\ell}_{n,\,m}(k)\,\mathcal{C}_{m,\,\ell}^{(\pm)}(k)=\delta_{n,\,0}\,
\frac{k}{\mathcal{S}_{0,\,\ell}(k)}, \qquad n=0, \, 1, \, \ldots
\,. \label{LJC}
\end{equation}
Here,
$\|J^{\ell}_{n,\,m}(k)\|=\|h^0_{n,\,m}-k^2\,A^{\ell}_{n,\,m}\|$ is
the so-called J-matrix. $\mathcal{S}_{n,\,\ell}$ are the
solutions of the system of equations
\begin{equation}
J^{\ell}_{n,\,m}(k)\,\mathcal{S}_{m,\,\ell}(k)=0, \qquad n=0, \,
1, \, \ldots, \label{LJS}
\end{equation}
\begin{equation}
\mathcal{S}_{n,\,\ell}(k)=\frac{\ell!\left(2\, \sin \zeta
\right)^{\ell+1}}{2\,(2\ell+1)!}(-\xi)^n\,{_2F_1}(-n, \ell+1;\;
2\ell+2;\;1-\xi^{-2}). \label{LS}
\end{equation}
It can easily be shown that the completeness relation for the
functions $\mathcal{S}_{n,\,\ell}$ of Ref. \cite{Broad} can be
rewritten as
\begin{equation}
\frac{2}{\pi}\int \limits _0^{\infty} dk\,
\mathcal{S}_{n,\,\ell}(k)\,A^{\ell}_{n',\,m}\,\mathcal{S}_{m,\,\ell}(k)=\delta_{n,
\,n'}.\label{LCRS}
\end{equation}

The coefficients $u_n(\mbox{i}\kappa_{\nu})$ of the expansion of
the bound state normalized wave function $\psi_{\nu}(r)$ with the
energy $-\kappa_{\nu}^2$ have the following asymptotic behaviour
\begin{equation}
u_n(\mbox{i}\kappa_{\nu})=w_n(\mbox{i}\kappa_{\nu}) \equiv
\mathcal{M}_{\nu}\,\mbox{i}^{\ell}\,
\mathcal{C}_{n,\,\ell}^{(+)}(\mbox{i}\kappa_{\nu}) \label{LBA}
\end{equation}
as $n\rightarrow\infty$.

Notice that the sine-like J-matrix solutions
$\widetilde{S}(r)=\sum \limits
_{n=0}^{\infty}\mathcal{S}_{n,\,\ell}(k)\,\phi_n^{\ell}(x)$ and
the cosine-like one $\widetilde{C}(r)=\sum \limits
_{n=0}^{\infty}\mathcal{C}_{n,\,\ell}(k)\,\phi_n^{\ell}(x)$, where
$\mathcal{C}_{n,\,\ell}(k)=\frac12\left(\mathcal{C}^{(+)}_{n,\,\ell}(k)+\right.$
$\left.\mathcal{C}^{(-)}_{n,\,\ell}(k) \right)$, have the
asymptotic behaviour (\ref{scasym}).

The the completeness relation (\ref{ComplSE}) is transformed into
\begin{equation}
\frac{2}{\pi}\int \limits _0^{\infty}u_n(k)\,
A^{\ell}_{n',\,m}\overline{u_m(k)}+\sum \limits
_{\nu}u_n(\mbox{i}\kappa_{\nu})\,A^{\ell}_{n',\,m}\,
\overline{u_m(\mbox{i}\kappa_{\nu})}=\delta_{n,\,n'}. \label{CR2}
\end{equation}

\subsection{Inverse problem}
In the framework of the J-matrix version \cite{ZK} of the inverse
scattering problem the spectral parameter set $\left\{\lambda_j,
\, Z_{N-1,\,j} \right\}_{j=0}^{N-1}$ is obtained from the
scattering data of the truncated Hamiltonian matrix of order $N$
in the orthogonal basis $\varphi_n^{\ell}=\sum_{m=0}^{N-1}
D_{n,\,m}^{\ell}\,\phi_m^{\ell}$, where
\begin{equation}\label{DM}
D_{n,\,m}^{\ell}=\left\{ \begin{array}{lr}
                   d_n^{\ell}, & n \ge m,\\[3mm]
                   0, & n<m, \\
                  \end{array} \right.
\quad d_n^{\ell} = \sqrt{\frac{2b\,n!}{(n+2\ell+2)!}},
\end{equation}
i. e.
\begin{equation}\label{bas1}
\varphi_n^{\ell}(x)=d_n^{\ell}(2br)^{\ell+1}e^{-br}L_n^{2\ell+2}(2br).
\end{equation}
Clearly the set $\left\{\lambda_j, \, Z_{N-1,\,j}
\right\}_{j=0}^{N-1}$ determines a tridiagonal Hamiltonian matrix
of order $N$ in any orthogonal basis
$\chi_n^{\ell}=\sum_{m=0}^{N-1}
P_{n,\,m}^{\ell}\,\varphi_m^{\ell}$ where $\| P_{n,\,m}^{\ell}\|$
is an arbitrary orthogonal $(N \times N)$-matrix of the form
\begin{equation}\label{PT}
\| P_{n,\,m}^{\ell}\|= \left(
\begin{array}{cccc}
  P_{0,\,0}^{\ell} & \cdots & P_{0,\,N-2}^{\ell} & 0 \\
  \vdots & \vdots & \vdots & \vdots \\
  P_{N-2,\,0}^{\ell} & \cdots & P_{N-2,\,N-2}^{\ell} & 0 \\
  0 & \cdots & 0 & 1
\end{array}
 \right).
\end{equation}
Let us assume that $\| P_{n,\,m}^{\ell}\|$ is the orthogonal
transformation matrix that performs the change from
$\left\{\varphi_n^{\ell}\right\}_{n=0}^{N-1}$ to the new basis
$\left\{\chi_n^{\ell}\right\}_{n=0}^{N-1}$ in which the kinetic
energy operator truncated matrix is tridiagonal. To perfect the
analogy with the oscillator basis case, let us denote the kinetic
energy operator $\frac{2\mu}{\hbar^2}H^0$ (\ref{H0}) tridiagonal
matrix in the basis $\left\{\chi_n^{\ell}\right\}_{n=0}^{N-1}$ by
$\|T_{n,\, m}^{\ell} \|$. The sought for Hamiltonian
$\frac{2\mu}{\hbar^2}H$ matrix $\|h_{n,\,m}\|$ of order $N$ is
presumed to be of a Jacobi form (\ref{hnm}) in the basis
$\left\{\chi_n^{\ell}\right\}_{n=0}^{N-1}$.

Thus, the first $N-1$ of the wave function $\psi(k, \, r)$
expansion  coefficients in the combined basis set
$\left\{\{\chi_n^{\ell}\}_{n=0}^{N-1},\;
\{\phi_n^{\ell}\}_{n=N}^{\infty}\right\}$ obey the equations
\begin{equation}\label{Eab}
 \begin{array}{c}
 a_0\,c_0(k)+b_1\,c_1(k)=k^2\,c_0(k)\\[3mm]
 b_n\,c_{n-1}(k)+a_n\,c_n(k)+b_{n+1}\,c_{n+1}(k)=k^2\,c_n(k),
 \quad n=1,\, \ldots \, N-2.\\
 \end{array}
\end{equation}
It is easy to check that a sufficient condition that the
algebraic version of the Marchenko method be applicable for the
construction of the tridiagonal Hamiltonian matrix (\ref{hnm}) is
that
\begin{equation}\label{NC}
a_{n+1}=T_{n+1,\,n+1}^{\ell},\; b_{n+1}=T_{n,\,n+1}^{\ell}, \quad
\mbox{for } n=M, \ldots,\, N-2, \; M=\lceil\frac{N}{2}\rceil.
\end{equation}
If $N$ is odd, to (\ref{NC}) must be added the constraint that
$a_M=T_{M,\,M}^{\ell}$. In this case $c_{M+1}=f_{M+1}$,
$c_{M}=f_{M}$, where $f_n$ satisfy the ``free'' equations
\begin{equation}\label{J1}
T_{n,n-1}^{\ell}f_{n-1}(k)+T_{n,n}^{\ell}f_{n}(k)
+T_{n,n+1}^{\ell}f_{n+1}(k)=k^2f_n(k),
  \quad n=1, \, \ldots, \, N-2,
\end{equation}
and we obtain for $n \le M-1$
\begin{equation}\label{EM}
c_n(k) = \sum \limits_{m=n}^{N-n-1}K_{n,\,m}f_{m}(k).
\end{equation}

Notice that in going from the initial Laguerre basis $\left\{
\phi_n^{\ell}\right\}_{n=0}^{\infty}$ to the combined basis set
$\left\{\{\chi_n^{\ell}\}_{n=0}^{N-1},\right.$ $\left.
\{\phi_n^{\ell}\}_{n=N}^{\infty}\right\}$ the submatrices
$\|h^0_{n,m}\|_{n,m=0}^{N-1}$ and \\
$\|A^{\ell}_{n,m}\|_{n,m=0}^{N-1}$ are transformed into
$\|T^{\ell}_{n,m}\|_{n,m=0}^{N-1}$ and the identity matrix of
order $N$ respectively. In addition, the elements $h^0_{N-1,N}$,
$A^{\ell}_{N-1,N}$ and $h^0_{N,N-1}$, $A^{\ell}_{N,N-1}$ are
multiplied by $d_{N-1}^{\ell}$. The rest of the (infinite)
matrices $\|h^0_{n,m}\|$ and $\|A^{\ell}_{n,m}\|$ is unaltered.
It thus follows that $f_n$ must satisfy (in addition to
(\ref{J1})) the equations
\begin{equation}\label{J2}
  T^{\ell}_{N-1,\,N-2}f_{N-2}(k)+
  T^{\ell}_{N-1,\,N-1}f_{N-1}(k)+
  d^{\ell}_{N-1}J^{\ell}_{N-1,\,N}f_N(k)=k^2f_{N-1}(k),
\end{equation}
\begin{equation}\label{J3}
  d^{\ell}_{N-1}J^{\ell}_{N,N-1}(k)f_{N-1}(k)+
  J^{\ell}_{N,N}(k)f_N(k)+
  J^{\ell}_{N,N+1}(k)f_{N+1}(k)=0.
\end{equation}
\begin{equation}\label{J4}
  J^{\ell}_{n,\,m}(k)f_m(k)=0, \quad n=N+1, \ldots\,.
\end{equation}
Putting $f_n=\mathcal{S}_{n,\ell}$ for $n \ge N$ and
$f_{N-1}=\mathcal{S}_{N-1,\ell}/d^{\ell}_{N-1}$ [in view of the
equation (\ref{J4}) and (\ref{J3}), respectively], from Eq.
(\ref{J2}) by using the tree-term recursion relation (\ref{J1}) we
obtain the coefficients $\widetilde{\mathcal{S}}_{n,\ell}$ with
$n=0,\ldots\,, N-2$. Similarly, setting
$f_n=\mathcal{C}^{(\pm)}_{n,\ell}$ for $n \ge N$ and
$f_{N-1}=\mathcal{C}^{(\pm)}_{N-1,\ell}/d^{\ell}_{N-1}$, we
obtain the coefficients
$\widetilde{\mathcal{C}}^{(\pm)}_{n,\ell}$ with $n=0,\ldots\,,
N-2$. From the Wronskian-like relation (see e.g. \cite{BR})
\begin{equation}\label{W1}
  J^{\ell}_{n+1,\,n}(k)\left(\mathcal{C}_{n+1,\ell}^{(\pm)}(k)\mathcal{S}_{n,\ell}(k)-
  \mathcal{C}_{n,\ell}^{(\pm)}(k)\mathcal{S}_{n+1,\ell}(k)\right)=k,
  \; n \ge 0
\end{equation}
it follows that
\begin{equation}\label{W2}
  T^{\ell}_{n+1,\,n}\left(\widetilde{\mathcal{C}}_{n+1,\ell}^{(\pm)}(k)
  \widetilde{\mathcal{S}}_{n,\ell}(k)-
  \widetilde{\mathcal{C}}_{n,\ell}^{(\pm)}(k)
  \widetilde{\mathcal{S}}_{n+1,\ell}(k)\right)=k, \; 0 \le n \le N-2.
\end{equation}
Besides, since the system of equations (\ref{LJS}) in
$\mathcal{S}_{n,\ell}$ is homogeneous, the sets
$\left\{\mathcal{S}_{n,\ell} \right\}_{n=0}^{\infty}$ and
$\left\{\widetilde{\mathcal{S}}_{n,\ell} \right\}_{n=0}^{\infty}$
are connected by a linear transformation and therefore
$\widetilde{\mathcal{S}}_{n,\ell}$ also satisfy the homogeneous
equation
\begin{equation}\label{S2}
  T^{\ell}_{0,0}\widetilde{\mathcal{S}}_{0,\ell}(k)+
  T^{\ell}_{0,1}\widetilde{\mathcal{S}}_{1,\ell}(k)=k^2
  \widetilde{\mathcal{S}}_{0,\ell}(k),
\end{equation}
whereas $\widetilde{\mathcal{C}}_{n,\ell}^{(\pm)}$ obey the
inhomogeneous one
\begin{equation}\label{C2}
  T^{\ell}_{0,0}\widetilde{\mathcal{C}}^{(\pm)}_{0,\ell}(k)+
  T^{\ell}_{0,1}\widetilde{\mathcal{C}}^{(\pm)}_{1,\ell}(k)=k^2
  \widetilde{\mathcal{C}}^{(\pm)}_{0,\ell}(k)+
  \frac{k}{\widetilde{\mathcal{S}}_{0,\ell}(k)}.
\end{equation}
Thus, the two sets, $\left\{\widetilde{\mathcal{S}}_{n,\ell}
\right\}_{n=0}^{\infty}$ and
$\left\{\widetilde{\mathcal{C}}_{n,\ell} \right\}_{n=0}^{\infty}$,
$\widetilde{\mathcal{C}}_{n,\ell}=\frac12\left(
\widetilde{\mathcal{C}}^{(-)}_{n,\ell}\right.+$
$\left.\widetilde{\mathcal{C}}^{(+)}_{n,\ell}\right)$, are
``free'' independent respectively sine-like
[$\widetilde{\mathcal{S}}_{n,\ell}= \mathcal{S}_{n,\ell}$, $n \ge
N$] and cosine-like [$\widetilde{\mathcal{C}}^{(\pm)}_{n,\ell}=
\mathcal{C}^{(\pm)}_{n,\ell}$, $n \ge N$] solutions to Eqs.
(\ref{J1})-(\ref{J4}) (see e.g. \cite{YF}).

From the above discussion it follows that to obtain $f_n$ with $0
\le n \le N-1$, which are involved in Eq. (\ref{EM}), we can
place $f_N=w_N$, $f_{N-1}=w_{N-1}/d^{\ell}_{N-1}$, where $w_n$
are defined by (\ref{Lkp}). Then, inserting this $f_{N}$ and
$f_{N-1}$ in Eq. (\ref{J2}) gives $f_{N-2}$. Once $f_{N-2}$ and
$f_{N-1}$ are known, $f_{n}$ for $n=N-3,\, \ldots,\, 0$ are
obtained by using the tree-term recursion relation (\ref{J1}).
$K_{n,\,m}$ are determined by the equations (\ref{KEq1}) and
(\ref{KEq2}) [in which the upper limit in the sums is equal to
$N-n-1$]. The expressions for $\{a_n\}$, $\{b_n\}$ are the same
as (\ref{abK}).

$Q_{n,\,n'}$ with $n \le N-1$ in Eqs. (\ref{KEq1}) and
(\ref{KEq2}) are defined [in view of the overlap matrix form in
the combined basis] by
\begin{equation}\label{CR3}
Q_{n,\,n'}=\frac{2}{\pi}\int \limits _0^{\infty}f_n(k)\,
\overline{f_{n'}(k)}+\sum \limits
_{\nu}f_n(\mbox{i}\kappa_{\nu})\,
\overline{f_{n'}(\mbox{i}\kappa_{\nu})}, \quad n' \le N-2,
\end{equation}
\begin{equation}\label{CR4}
 \begin{array}{c}
Q_{n,\,N-1}=\frac{2}{\pi}\int \limits
_0^{\infty}f_n(k)\,[\overline{w_{N-1}(k)}/d^{\ell}_{N-1}+
d^{\ell}_{N-1}A^{\ell}_{N-1,N}\overline{w_{N}(k)}]+
\qquad\qquad \qquad \\[3mm]
\qquad +\sum \limits _{\nu}f_n(\mbox{i}\kappa_{\nu})\,[
\overline{w_{N-1}(\mbox{i}\kappa_{\nu})}/d^{\ell}_{N-1}+
d^{\ell}_{N-1}A^{\ell}_{N-1,N}\overline{w_{N}
(\mbox{i}\kappa_{\nu})}],\\
 \end{array}
\end{equation}
\begin{equation}\label{CR5}
Q_{n,\,n'}=\frac{2}{\pi}\int \limits
_0^{\infty}f_n(k)\,A^{\ell}_{n',m}\overline{w_m(k)}+ \sum \limits
_{\nu}f_n(\mbox{i}\kappa_{\nu})\,A^{\ell}_{n',m}
\overline{w_m(\mbox{i}\kappa_{\nu})}, \quad n' \ge N.
\end{equation}
Notice that at large $k$, as seen in Eq. (\ref{LCpm}),
$\mathcal{C}^{(\pm)}_{n,\ell}(k) \sim k^{\ell}$. Thus, as in the
case of the oscillator basis, we should restrict ourselves to the
description of the scattering data on a finite energy interval,
beyond the boundary of which the phase shift needs generally to be
modified to provide the convergence of the integrals in Eqs.
(\ref{CR3})-(\ref{CR5}).

\section{Conclusion}
In the potential scattering case the finite-difference approach and J-matrix method
share the tridiagonal representation of the Hamiltonian. The analogy can be carried over
to the inverse scattering problem formalism. Here, the J-matrix version of the Marchenko
equation algebraic analogue is formulated and its numerical realization features are
considered. The merit of JME is that it is free from a parameter fit inherent in the
previous J-matrix inverse scattering approach \cite{Z1,Z2,ZK,PC}. We also construct a
tridiagonal Hamiltinian matrix of some order $M$ in an orthogonalized Laguerre basis; in
doing so it is sufficient to tridiagonalize the matrix representation of the reference
Hamiltonian $H^0$ in the finite orthogonal basis subset of size $N=2M$. As has been shown
in Ref. \cite{PC}, in the two coupled-channel case without threshold the sought-for
interaction matrix may be of a ``quasi-tridiagonal'' form. On this assumption JME can be
easily expanded to multichannel scattering.

\subsection*{Acknowledgments}
The author acknowledge helpful conversations with A~.M.~Shirokov.
This work was partially supported by the State Program ``Russian
Universities'', by the Russian Foundation of Basic Research grant
No 02-02-17316.

\newpage
\begin{figure}
\centerline{\psfig{figure=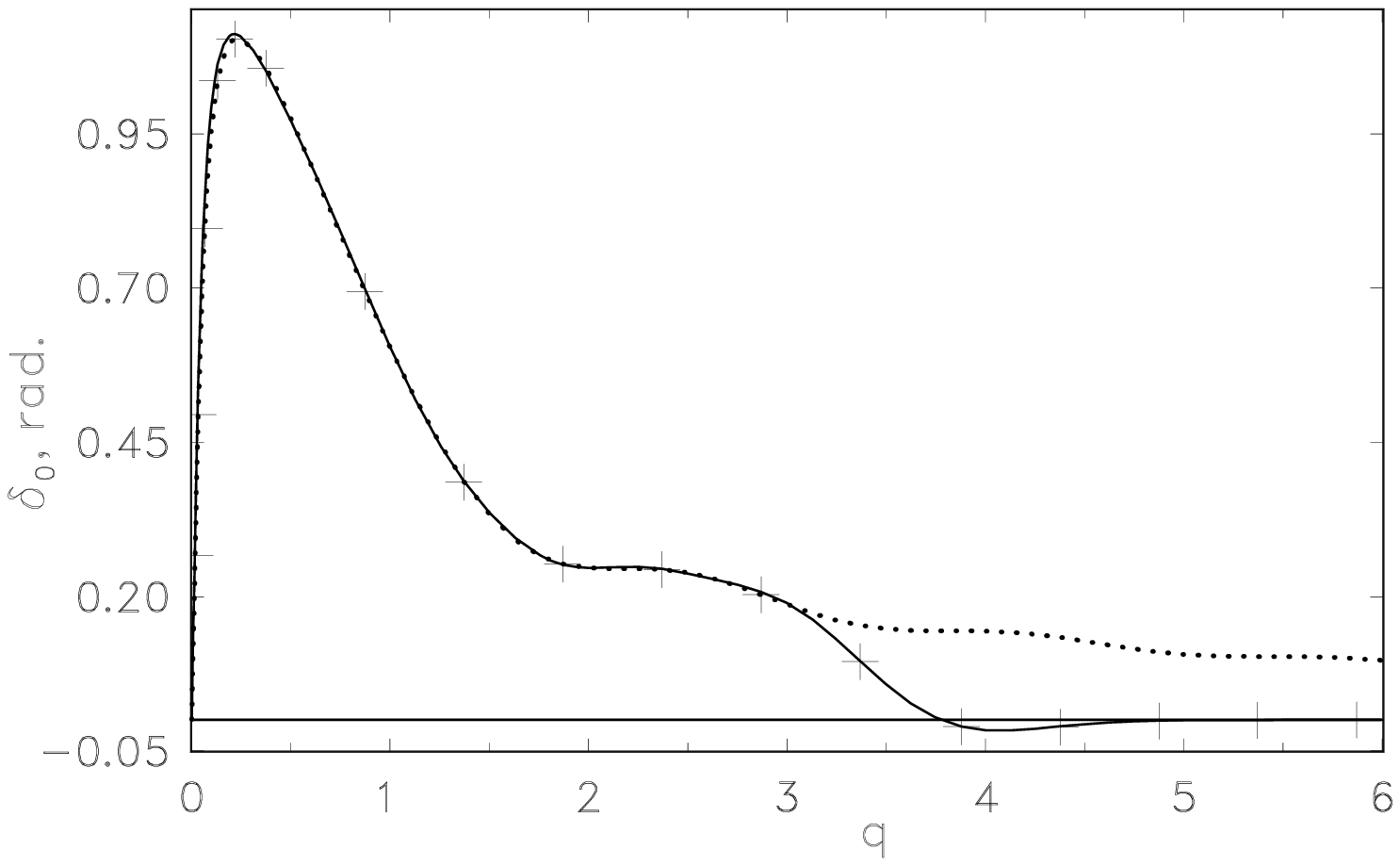,width=1.3\textwidth}}\caption{}
 \label{fig1}
\end{figure}
\clearpage

\newpage
\begin{figure}
\centerline{\psfig{figure=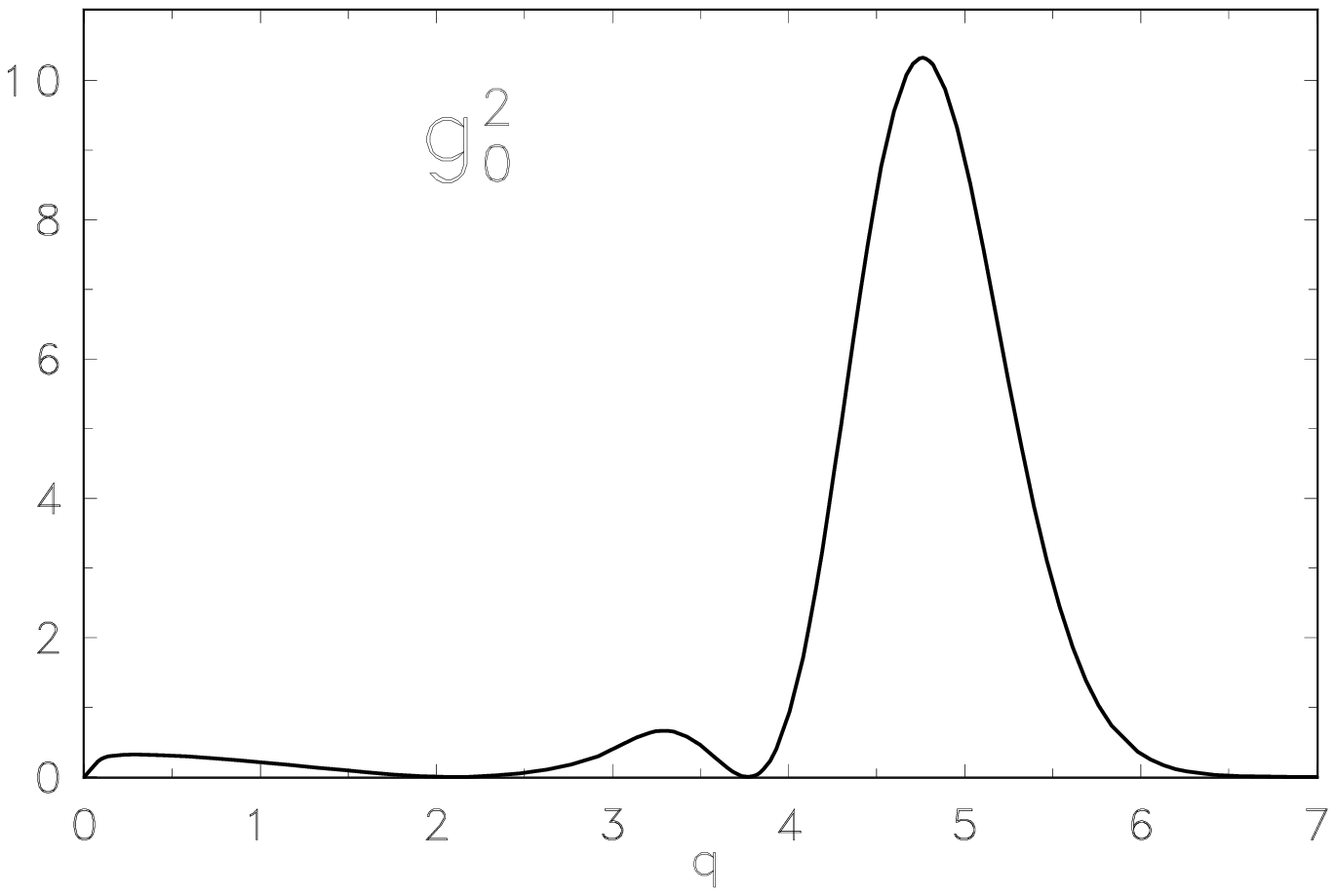,width=1.3\textwidth}}\caption{}
 \label{fig2}
\end{figure}
\clearpage

\newpage
\begin{figure}
\centerline{\psfig{figure=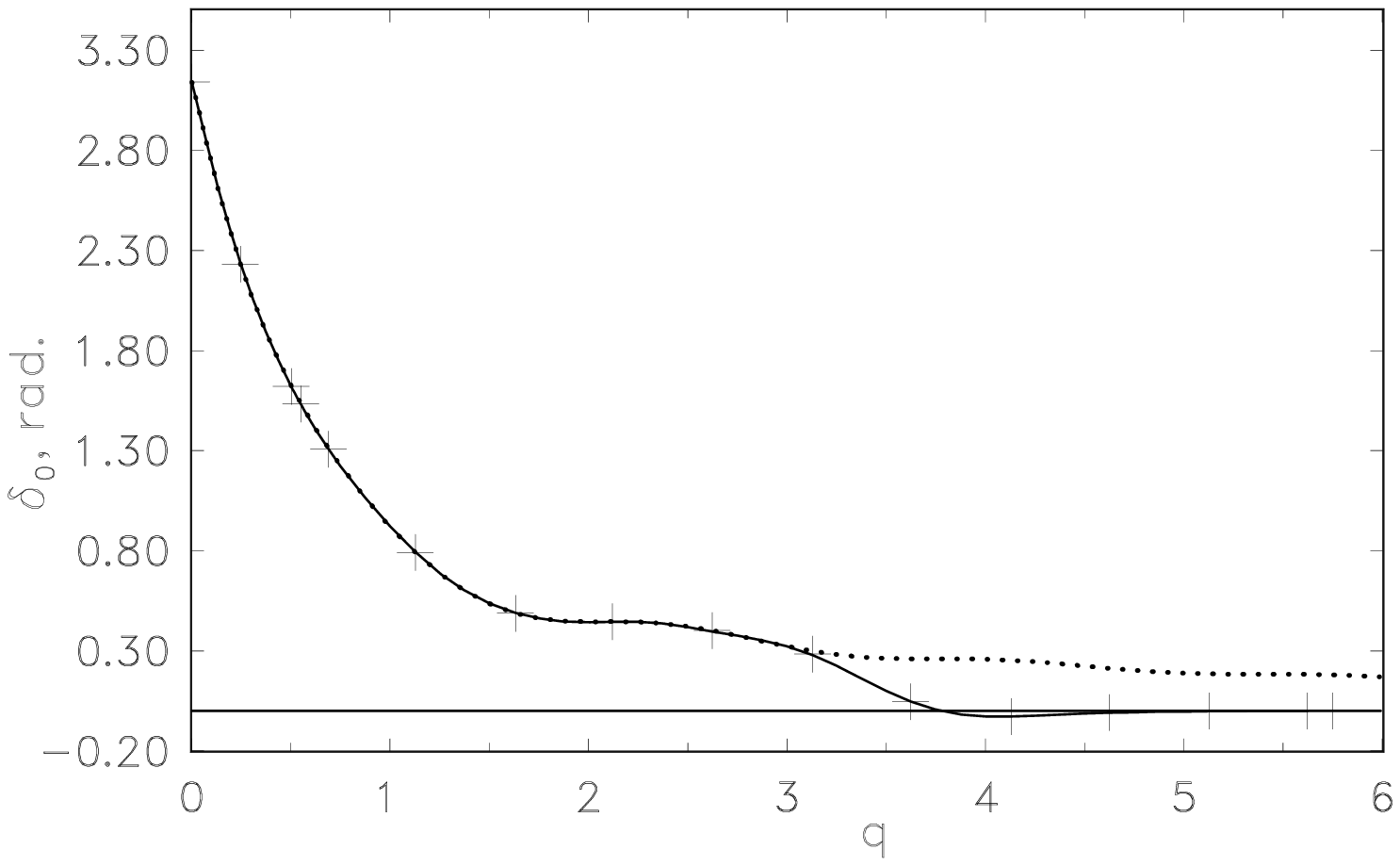,width=1.3\textwidth}}\caption{}
 \label{fig3}
\end{figure}
\clearpage

\newpage
\begin{center}
Table
\end{center}
$$
\begin{array}{ccc|cc}
\hline \hline \multicolumn{5}{c}{\vphantom{{{C^C}_C}^C} N=7, \; \rho=\frac{R}{2}}\\
\hline
\multicolumn{3}{c}{\begin{array}{c}\kappa R=.6512647458, \\
\mathcal{M}R^{1/2}=1.6017576599\\
\end{array}
} &
 \multicolumn{2}{|c}{\begin{array}{c}\kappa R=.6380449999, \\
\mathcal{M}R^{1/2}=1.5833238674\\
\end{array}
}\\
\hline
j & Z_{N-1, \, j} & \lambda_j & Z_{N-1, \, j} & \lambda_j\\
\hline
  \begin{array}{c}
   0\\ 1\\ 2\\ 3\\ 4\\ 5\\ 6\\
  \end{array}
 & \begin{array}{l}
0.0356514517\\ 0.1482712147\\ 0.2309801539\\ 0.3094585382\\ 0.4084394267\\
0.5275945630\\ 0.6184249465\\
   \end{array}
 & \begin{array}{l}
-0.0381260178\\ \phantom{-} 0.4605384282\\ \phantom{-} 1.3246452781\\
\phantom{-} 2.5044702689\\ \phantom{-} 4.1865934000\\ \phantom{-} 6.7063360348\\
\phantom{-} 10.1425219887\\
\end{array}
 & \begin{array}{l}
0.0362075259\\ 0.1482712147\\ 0.2309801539\\ 0.3094585382\\ 0.4084394267\\
0.5275945630\\ 0.6183926387\\
   \end{array}
 & \begin{array}{l}
-0.0353279575\\ \phantom{-} 0.4605384282\\ \phantom{-} 1.3246452781\\
\phantom{-} 2.5044702689\\ \phantom{-} 4.1865934000\\ \phantom{-} 6.7063360348\\
\phantom{-} 10.1425219887\\
\end{array}\\
\hline \hline
\end{array}
$$


\begin{thebibliography}{99}

\bibitem{HY}
 E.~J.~Heller, H.~A.~Yamani, Phys. Rev. A {\bf 9}, 1209 (1974).

\bibitem{Z1}
 S.~A.~Zaitsev, Teoret. Mat. Fiz. {\bf 115}, 263 (1998) [Theor.
 Math. Phys. {\bf 115}, 575 (1998)].

\bibitem{Z2}
 S.~A.~Zaitsev, Teoret, Mat. Fiz. {\bf 121}, 424 (1999) [Theor. Math. Phys.
 {\bf 121}, 1617 (1999)].

\bibitem{ZK}
 S.~A.~Zaitsev, E.~I.~Kramar, J. Phys. G, {\bf 27}, 2037 (2001).

\bibitem{PC}
 A.~M.~Shirokov, A.~I.~Mazur,  S.~A.~Zaytsev, J.~P.~Vary, T.~A.~Weber,
 Phys. Rev. C {\bf 70} (2004) 044005.

\bibitem{YAA}
 H.~A.~Yamani, A.~D.Alhaidari, and M.~S.~Abdelmonem, Phys. Rev.
 A {\bf 64}, 042703 (2001).

\bibitem{YF}
 H.~A.~Yamani, L.~Fishman, J. Math. Phys. {\bf 16}, 410 (1975).

\bibitem{G}
 K.~Ghanbari, IP {\bf 17}, 211 (2001).

\bibitem{GW}
 G.~M.~L.~Gladwell, N.~B.~Willms, IP {\bf 5}, 165 (1989).

\bibitem{Chabanov}
 V.~M.~Chabanov, J. Phys. A, {\bf 37}, 9139 (2004).

\bibitem{Case}
 K.~M.~Case, J. Math. Phys. {\bf 14}, 916 (1973)

\bibitem{Zakhariev}
 B.~N.~Zakhariev, A.~A.~Suzko, {\sl Direct and inverse problems. In: Potentials in
 quantum scattering} (2-nd ed. Berlin, Heidelberg, New York: Springer-Verlag,
 1990).

\bibitem{Baz}
A.~I.~Baz, Ya.~B.~Zeldovitch, and A.~M.~Perelomov, {\sl
Scattering, Reactions and Decays in Non-relativistic Quantum
Mechanics} (Moscow: Nauka, 1971).

\bibitem{CS}
K.~Chadan, P.~C.~Sabatier, {\sl Inverse Problems in Quantum
Scattering Theory} (New York, Heidelberg, Berlin:
Springer-Verlag, 1977).

\bibitem{Broad}
 J.~T.~Broad, Phys. Rev. A {\bf 31}, 1494 (1985).

\bibitem{BR}
 J.~T.~Broad, W.~P.~Reinhardt, J. Phys. B {\bf 9}, 1491 (1976).

\end{thebibliography}
\end{document}